\documentclass[vecphys]{svmult}

\usepackage{makeidx}         
\usepackage{graphicx}        
\usepackage{multicol}        
\usepackage[bottom]{footmisc}

\makeindex             

\begin{document}

\title*{ Self assembly of a model multicellular organism resembling the Dictyostelium slime molds}
\titlerunning{Self assembling slime mold}
\author{Graeme J. Ackland\inst{1}, Richard Hanes\inst{1} \and Morrel H. Cohen
\inst{2}}
\institute{$^1$ School of Physics, SUPA, The University of Edinburgh, Mayfield Road, 
Edinburgh EH9 3JZ, UK 
\texttt{ gjackland@ed.ac.uk};  $^2$ Department of Physics and Astronomy, Rutgers University, 136 Frelinghuysen Road, 
Piscataway NJ 08854 USA and Department of Chemistry, Princeton University, Princeton NJ 08544 USA
\texttt{mhcohen@prodigy.net}}

%
%
\maketitle

\section{Abstract}

The evolution of multicellular organisms from monocellular ancestors
represents one of the greatest advances of the history of life.  The
assembly of such multicellular organisms requires signalling and
response between cells: over millions of years these signalling
processes have become extremely sophisticated and refined by
evolution, such that study of modern organisms may not be able to shed
much light on the original ancient processes .  Here we are interested
in determining how simple a signalling method can be, while still
achieving self-assembly. In 2D a coupled cellular automaton/differential equation approach models organisms and chemotaxic chemicals, producing spiralling
aggregation.  In 3D Lennard-Jones-like particles are used to
represent single cells, and their evolution in response to signalling
is followed by molecular dynamics. It is found that if a single cell
is able to emit a signal which induces others to move towards it, then
a colony of single-cell organisms can assemble into shapes as complex
as a tower, a ball atop a stalk, or a fast-moving slug.  The
similarity with the behaviour of modern Dictyostelium slime molds
signalling with cyclic adenosine monophosphate (cAMP) is striking.

\section{Introduction} 
\label{sec:2}

The myriad shapes and complex adaptations
which are observed in modern organisms may suggest that evolution is
unbounded in its possibilities: yet several "simple" machines,
e.g. the wheel and the double pulley have no counterpart in the
natural world, while the complexity of some natural systems, such as
the eye or the brain, remain a source of amazement.  Thus it is clear
that human intuition about what is "simple" is different from what is
evolutionarily achievable - i.e. what can be built incrementally
rather than by design.

The transition from unicellular to multicellular lifeforms is one of
the most dramatic changes in evolutionary history (see e.g. Bonner, 1997 and references therein)\cite{bonner}.  
The exact trigger
for cooperative behaviour is unknown; however, some form of signalling
between cells must have been involved.  In this paper a model is
introduced for investigating by simulation the types of multicellular
shapes which could emerge in response to simple signalling.

The emphasis in this work is to explore how simple a morphogenetic
system can be that will lead to self assembly, what the mechanism of
assembly is, and what morphologies might result.  In particular, it
does not attempt to model any particular modern organism, and, where
choices have to be made, the simplest option is preferred.

\begin{figure}[htb]
\centerline{
\includegraphics[height=10cm]{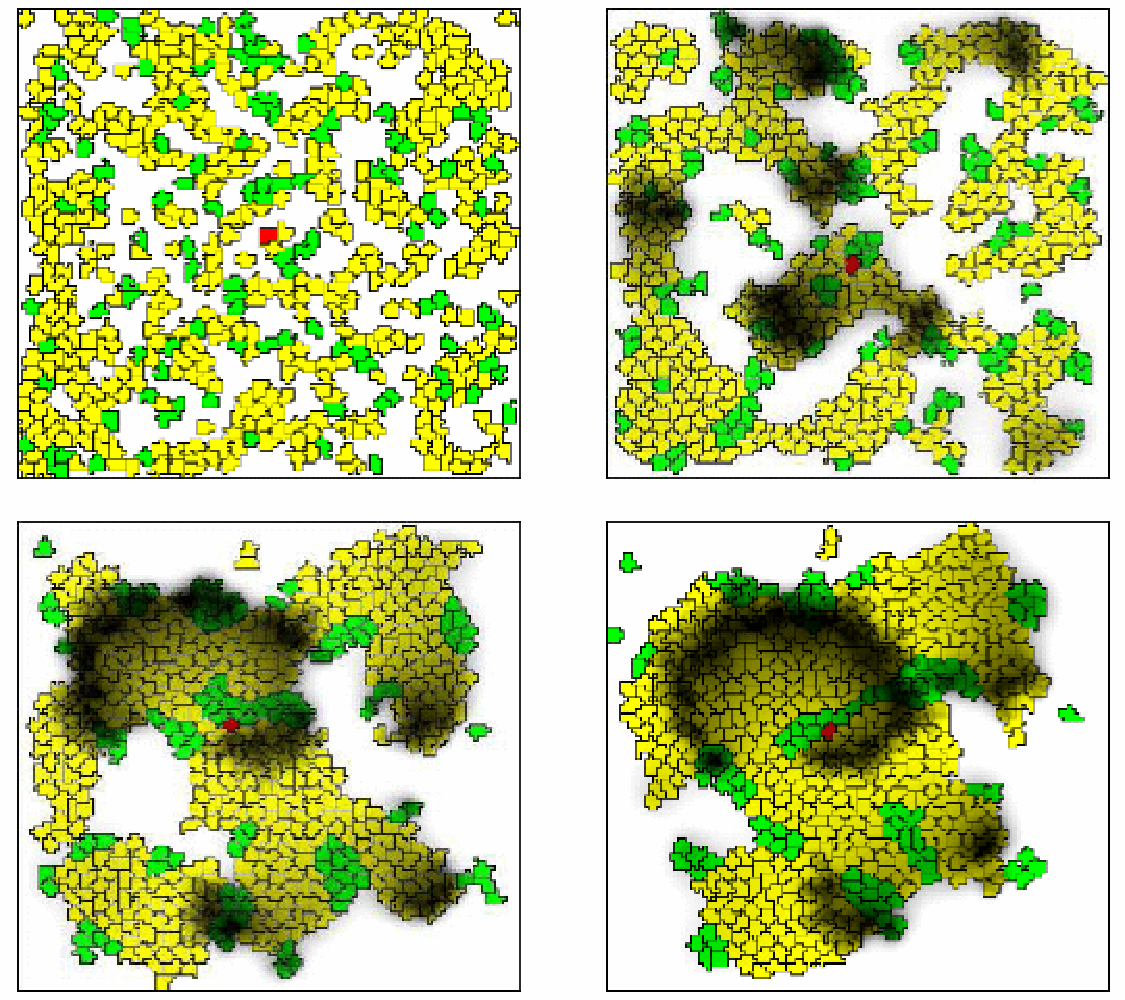}}
        \caption{ The formation of spiral waves of cAMP by a autocycler. 
The red amoeba is auto-cycling, green are pre-stalk and
yellow are pre-spore. Dark shading shows the cAMP concentration.
The images show that the different amoeba
types sort themselves into groups during aggregation. (a) After
Initialisation,
(b) 500 iterations, (c) 1,000 iterations, (d) 3,000 iterations. 
\label{stimspiral}       
}
\end{figure}

\section{Accumulation, Spirals and Self-organisation in 2D}

The accumulation of Dictyostelium Discoidium (Dd) in the early stages
of aggregation can be modelled using a two stage Cellular Automata
(CA) /Differential equation (DE) model similar to that suggested by
Hogeweg\cite{Hog}. A detailed description of our implementation,
coding and an applet can be found online at www.ph.ed.ac.uk/nania/dicty/dicty.html.

The CA model is based on work by Glazier and Graner\cite{gg}.  On a 2D
square lattice each amoeba covers multiple sites (with ideal value
$\nu$) and is assigned a unique identification number $\sigma$.  Using multiple sites is essential to allow the amoeba to pass each other and apply pressure.
The three types of amoeba (autocycling(a), prestalk(k), prespore(p)) 
can each  be in one of three states (Ready (s=0), Excited(s=1), Refractory(s=-1)).
Amoeba are initialised with area $\nu$ at random positions on the 
lattice\cite{Hanes}.  Each lattice site $i$ is assigned an energy

\begin{equation}\label{CAEn}
H_i = \sum_j J(\tau_{ij}) + \lambda(\nu-V(\sigma_i)) 
\end{equation}  

where the sum runs over eight nearest neighbours.  J($\tau$) is given
in table \ref{tautable} and $V( \sigma_i))$ is the number of sites
occupied by the amoeba at $i$ (labelled $\sigma_i$).

\begin{table}
\centering
\caption{Adhesion energies.  Other parameter used in the simulations are:
CA Size 120x120; 
Number of Amoeba 540; 
$\nu$=16;
Mobility T=2;
Membrane Elasticity $\lambda=1$ 
Chemotaxis Constant $\mu=30$ D=0.1 $c_t=0.2$ $\gamma=0.04$}
\label{tab:1}       
\begin{tabular}{lllll}
\hline\noalign{\smallskip}
$\tau_{ij}$ & 0 & a & k & p \\
\noalign{\smallskip}\hline\noalign{\smallskip}
0 &0 &2 &2 &2 \\ 
a &2 &3 &4 &4 \\ 
k &2 &4 &3 &4 \\ 
p &2 &4 &4 &4 \\
\noalign{\smallskip}\hline
\label{tautable} \end{tabular}
\end{table}

The CA is updated by copying the $\sigma_j$ onto a neighbouring site $i$ 
with a  probability according to the Metropolis algorithm\cite{metro}. 
An additional term, $\Delta H_{ij} = \mu{c_i-c_j}$, described below, is 
added for Excited amoeba in the presence of cAMP.
The ``temperature'' parameter T controls the mobility of the amoeba. 

The CA alone exhibits cell sorting behaviour\cite{gg,Hanes} due to
differential adhesion of different cell types\cite{steinberg}.

Our model varies from previous work\cite{Hog} in that the signalling 
chemical cAMP is added directly by the amoeba when they are in 
the excited state, and the time advancement is controlled by the 
diffusion equation with 
a term ($\gamma$)to account for the breakdown of cAMP by phosphodiesterase. 
Thus the concentration of cAMP $c(x,y)$ is given by:

\begin{equation}
\frac{dc(x,y)}{dt} = D\nabla^2c(x,y) - \gamma c(x,y) + \delta(s(x,y)-1),
\label{camp}
\end{equation}

where $s(x,y)$ is the CA state of the amoeba at position $i\equiv(x,y)$.

One complete iteration of the cAMP field takes place for each iteration 
of the CA. 

Dynamics at the amoeba level cycle between the three states: R
transforms to E once cAMP concentration integrated over the amoeba
exceeds some threshhold $c>c_t$.  The amoeba remains E for 100s, then
becomes F for 500s before returning to R.

Autocycling amoeba are different. They simply emit cAMP once every
600s, and remain F.  There is no evidence that such autocycling amoeba
exist in nature, and we shall see later that they are unnecessary for
explaining the observed dynamics.

\begin{figure}[htb]
\centerline{
\includegraphics[height=10cm]{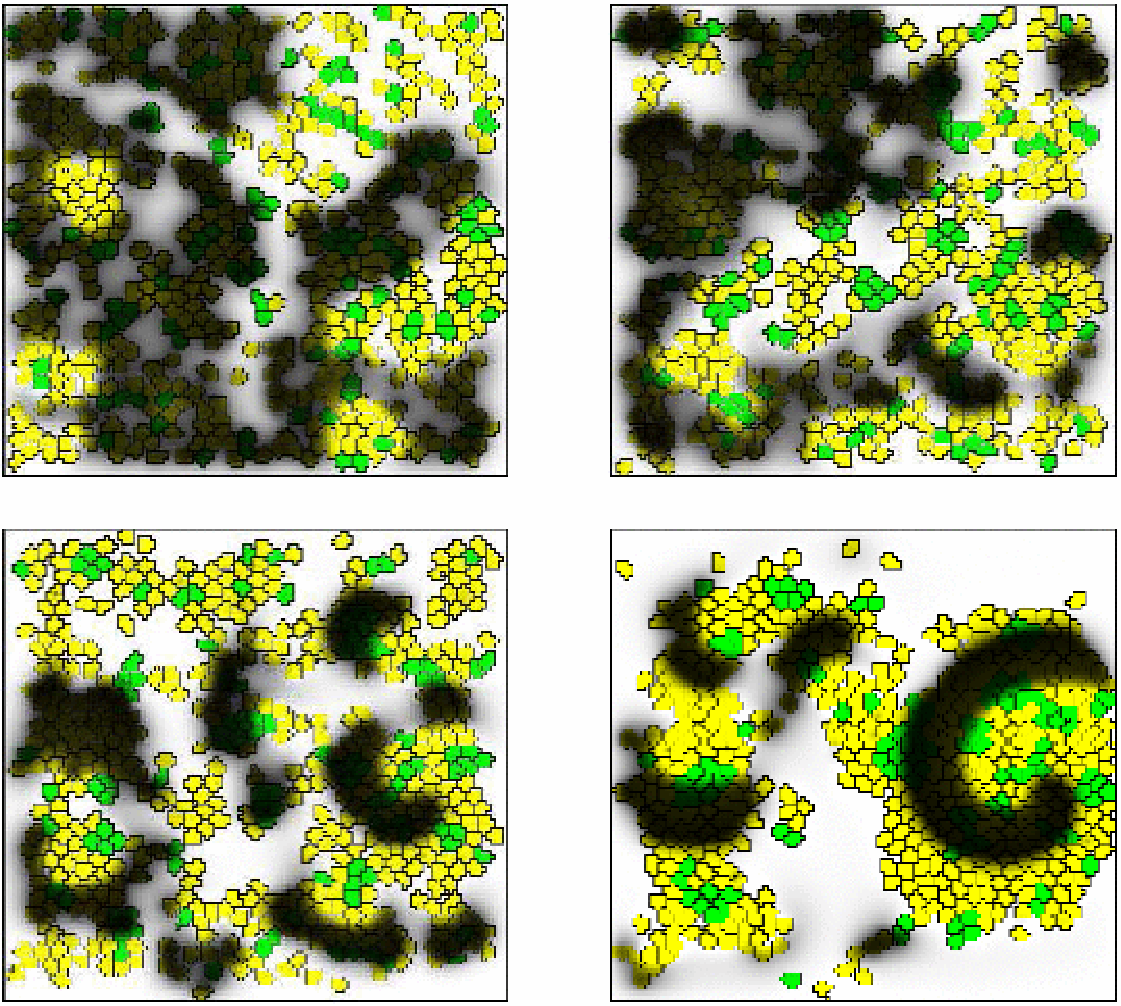}}
        \caption{ As in Fig.\ref{stimspiral} without the autocycler. 
(a) After
Initialisation,
(b) 500 iterations, (c) 1,000 iterations, (d) 3,000 iterations. 
\label{sponspiral}       
}
\end{figure}
\subsection{Spiral Formation}

With one autocycler amoeba, 108 pre-stalk and 431 pre-spore cells
spiral patterns are formed (Fig.\ref{stimspiral}). Aggregation happens
quickly, followed by the generation of a spiral wave initiated by the
autocycler.  The spiral wave causes the aggregating amoeba to form
streams, which coalesce into a single large stream.  This is broadly similar to
previous work\cite{Hog}.  Cell sorting is a slower process, but can be
seen in (Fig.\ref{stimspiral}).  Gaps between amoeba are important: if
the amoeba density is increased to fill all space, no such spirals are
formed.

Without the autocycler, the spiral formation can be triggered by
occasional, random emissions of cAMP from randomly chosen amoeba. The
physical reasoning is that when the amoeba begin to starve, they all
emit small amounts of cAMP spontaneously, with no special triggering
amoeba to start the process.  The results show that initially the
noise creates disorder among the amoebas, but after a small amount of
time, if a cAMP waves travels around a closed loop, a self sustaining
spiral is formed which then controls the aggregation process,
Fig.\ref{sponspiral} This spontaneous formation takes rather longer
than the stimulated equivalent, but is equally robust once established.

\subsection{Streaming}

The formation of streams during aggregation is an interesting property
of Dd amoeba.
As the cAMP wave propagates along the stream, it travels faster down
the centre as the amoeba density is higher in the centre than at the
edges\cite{vonOss}.  This creates a curvature in the wavefront, causing the
chemotactically moving amoeba to push towards the centre of the stream, 
as each wave passes. Diffusion from the edges of
the stream excites nearby amoeba which see the strongest gradient of
cAMP towards the stream and join. The overall effect is that
waves of cAMP make small groups of amoeba form small streams, which
attract more amoeba, and the small streams join together to create
bigger streams which flow towards the aggregation centre.  A
further observation is that if the centre of aggregation is a self
sustaining spiral, the streams spiral into the aggregation centre,
which appears to rotate. This can be seen in Figure \ref{stimspiral}, where the
spiral is rotating in a clockwise direction and the aggregation centre
is rotating in the anti-clockwise direction. Figure \ref{sponspiral} shows
aggregation which has been controlled by the circular waves of cAMP
spreading from the auto-cycling amoeba, and
in this experiment, the aggregation centre does not appear to be rotating.

\begin{figure}[htb]
\centerline{
\includegraphics[height=6cm]{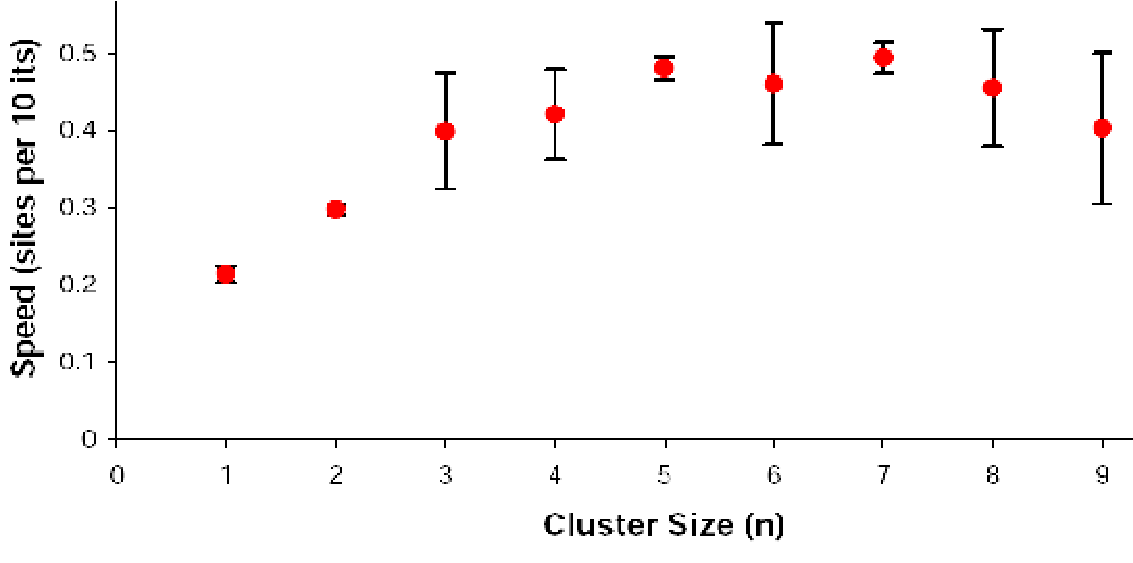}}
        \caption{ The speed of a group of amoebas vs number in the group. Results 
were averaged over five runs and the standard deviation displayed in the figure
\label{race}       
}
\end{figure}

\subsection{Clumping}

When the amoebas' resource supply becomes scarce, they need to move to
an area with more resources as quickly as they can in order to
survive. By grouping together, amoeba travel more quickly than if they
move independently. Bonner showed by experiment that this happens in
the migrating slug [6].  Figure.\ref{race} shows the results of  a synthetic experiment to
determine how the velocity of a group of amoebas changes with the
number of amoeba in the group.  We applied a fixed external cAMP gradient and released groups of different sizes to move through chemotaxis.
In this experiment, an individual amoeba travels less than half the 
speed of a larger group, and that for a large n, the speed tends to a constant, which is in 
accordance with observed experimental results\cite{Inouye}.

\begin{figure}[htb]
\centerline{
\includegraphics[height=6cm,angle=90]{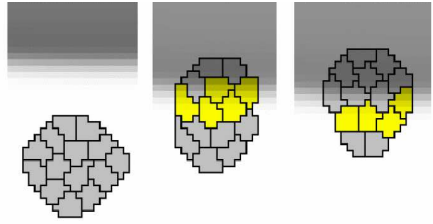}}
        \caption{Three images showing elongation. A wave of cAMP (grey) passes 
through a group of amoeba. 
The grey amoeba are Ready or Refractory and the yellow are Excited. The wave will have passed before the amoeba cease to be Refractory.
\label{elongate}       
}
\end{figure}

There are several reasons why amoeba move faster in a group, the most
dominant effect comes from adhesion between cell membranes. If an
amoeba moves independently, it can travel a short distance during its
excited period. If this amoeba is surrounded by other amoebas, it gets
pulled along by the amoeba moving before it and pushed by the amoeba
moving after it, travelling further for each wave of cAMP.  This
experiment used a static cAMP field to demonstrate the how the pushing
and pulling motion and adhesion allow amoeba to travel faster in a
group. Other effects arise when waves of cAMP are
used. Figure\ref{elongate} shows the motion of a group of amoeba when
a wave of cAMP is passed over them. With no cAMP present, they adopt
circular shape to minimise the boundary. As a wave of cAMP passes
through, the group elongates as the amoeba at the front move up the
wave front.  This elongation makes the amoeba at the rear of the group
move forward to minimise the surface once again.

\section{Forming a slug in 3D} 
\label{sec:3}
To investigate slug formation we coarse-grain the model described above, 
representing the amoeba as point particles and the cAMP field as an attraction to the pacemaker. This model becomes essentially a molecular dynamics calculation.  These are serious simplifications, but as we shall see this 
model captures the relevant dynamic instabilities which cause particles to form a tower.

The interactions in the model are gravity, the
reaction force of the ground, 3-dimensional stochastic noise, viscous
damping, interacting amoebas of finite size, and the response to
signalling.  Gravity provides a constant downward force, and the
ground reaction force is perfectly inelastic (if any amoeba attempts to
move below z=0, its velocity is set to zero and position to z=0).  The
stochastic force is drawn randomly from a uniform distribution at each
timestep, the damping force is proportional to velocity, and in the
opposite direction.  The amoebas are described by a Lennard-Jones
11-5 potential, which gives a separation-dependent force with a short-range
repulsion and a long-range attraction.  This determines the size of
the amoebas, makes them slightly sticky when they approach one another
and implicitly gives them spherical symmetry.  In keeping with our
fundamental approach, we regard this as the simplest assumption which
is still reasonable. It would certainly be possible to use elliptical
amoebas (or some other shape) or a deformable membrane containing
incompressible fluid.  If our aim were simulation of a specific
organism this would be appropriate; however in the hope of obtaining
generic trends, we prefer the simpler approach.  

The signalling
between amoebas and consequent self-propulsion is similarly simple.  One
amoeba is assumed to start signalling, and all others move towards it. 
The assumption is that each individual amoeba is
motile, either dragging itself along any available surface or
propelling itself with flagellae. The specific mechanism is
unimportant to the general argument, but the motion is initially
confined to the xy direction since the amoebas cannot fly. For states
where all the amoebas are packed together (all those investigated here)
any self-propulsion upward in the z direction would be either
impossible (for a surface amoeba) or counterbalanced by the downward
pull of similarly self-propelled amoebas below.  More complex
signalling, such as the relayed cAMP chemicals emitted by slime molds,
has been investigated by previously, producing realistic simulations
of actual organisms.  Here, again, we seek to find the structures
realised by the simplest possible signalling.  

 The equations of
motion are integrated using a velocity verlet algorithm.

\begin{figure}[htb]
\centerline{
\includegraphics[height=10cm]{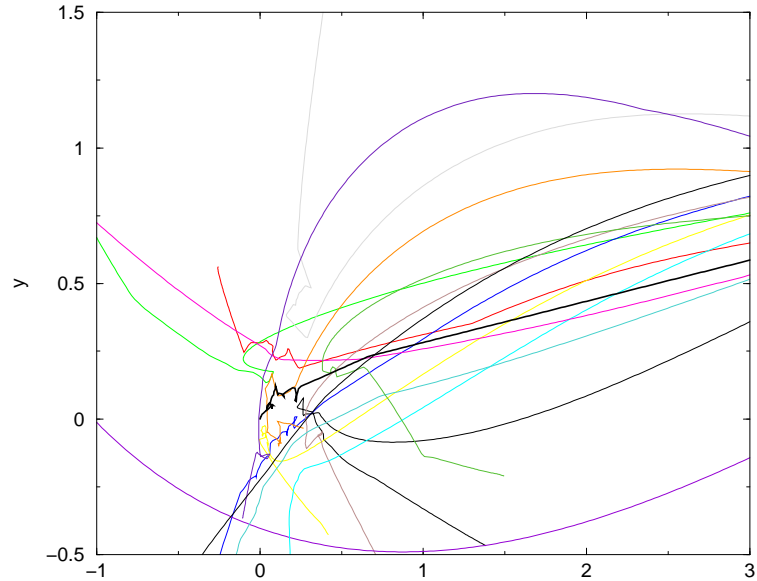}}
        \caption{
 Typical trajectories of some of the 3000 amoebas projected
onto the surface plane.  Only a few of the amoebas are shown here: if all
were plotting the "jiggling" would be lost and the appearance would be
of inward radial streaming motion, followed by coordinated movement to
the right.  Amoebas begin in the vicinity of the origin and
jiggle around while the tower assembles.  Some amoebas (see lowest
trajectory) do not form the tower, rather they provide the pressure
which causes the uplifting instability to create the tower, then join the coordinated motion once it
is established. Once the tower falls, rapid cooperative motion of the
multiamoeba slug takes the particles out of this displayed region to
the right hand side: the smoothness of the trajectories within the
slug arises from their velocities being much higher than the effect of
the stochastic noise, and their being constrained by neighbours.  
\label{fig:1}       
}
\end{figure}

The
mechanism by which the original amoeba starts to signal is unimportant: 
in particular we do not propose that there is anything special
about it except that it changes to the signalling state.  For example,
it could even be the first amoeba to die of starvation, and the "signal"
could be the decay products of decomposition.  As we shall see, there
is nothing in the subsequent behaviour which confers any particular
advantage on the signalling amoeba.

Specifically, the force on each amoeba $i$ is:

\begin{equation}
 {\bf F_i} = {\bf F_{sig}} +  \sum_j e_0 
\left [ \frac{a_o}{r_{ij}}^12 - \frac{a_o}{r_{ij}}^6 \right ]
 \frac{\bf{r_{ij}}}{r_{ij}}
  +  m{\bf g}  - d {\bf v_i} + {\bf F_{rand}}
\end{equation}

The observed behaviour depends, of course, on the
strengths of the various interactions and the number of amoebas.  A
basic reference shape is a low dome or a sphere, which would be
adopted in the absence of signalling by particles interacting via the
potential, gravity and stochastic noise alone.  These shapes
are typical of those adopted by liquid drops, determined by a balance
between surface tension and gravity.

\begin{figure}[htb]
\centerline{
\includegraphics[height=8cm]{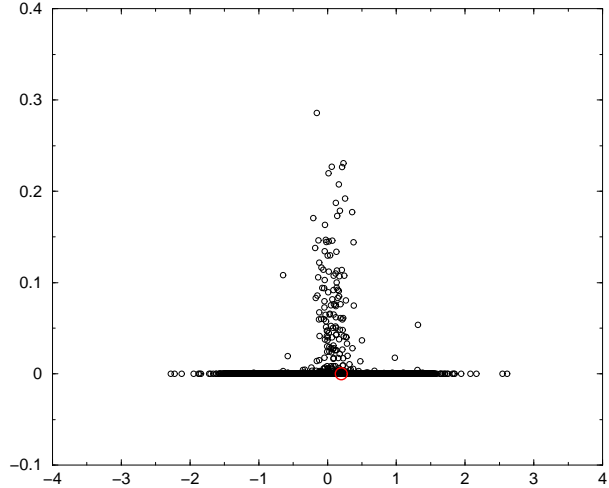}}
        \caption{
 Initial instability for 3000 amoebas, axes are horizontal and
vertical respectively, individual amoebas are shown by circles, the
radius of which is arbitrary.  The red circle shows the position the
signalling amoeba: although the signaller causes the instability it is
not, in this case, among the first few amoebas to be lifted up. Typical
parameter values are $F_{sig}$= 10, $F_{rand}$=0.04, $mg$ = 0.001 $d$
= 100 $a_0$=0.02 $e_0$=30, with an integration timestep of $10^{-5}$.
\label{fig:2}       
}
\end{figure}

\begin{figure}[htb]
\centerline{
\includegraphics[height=6cm]{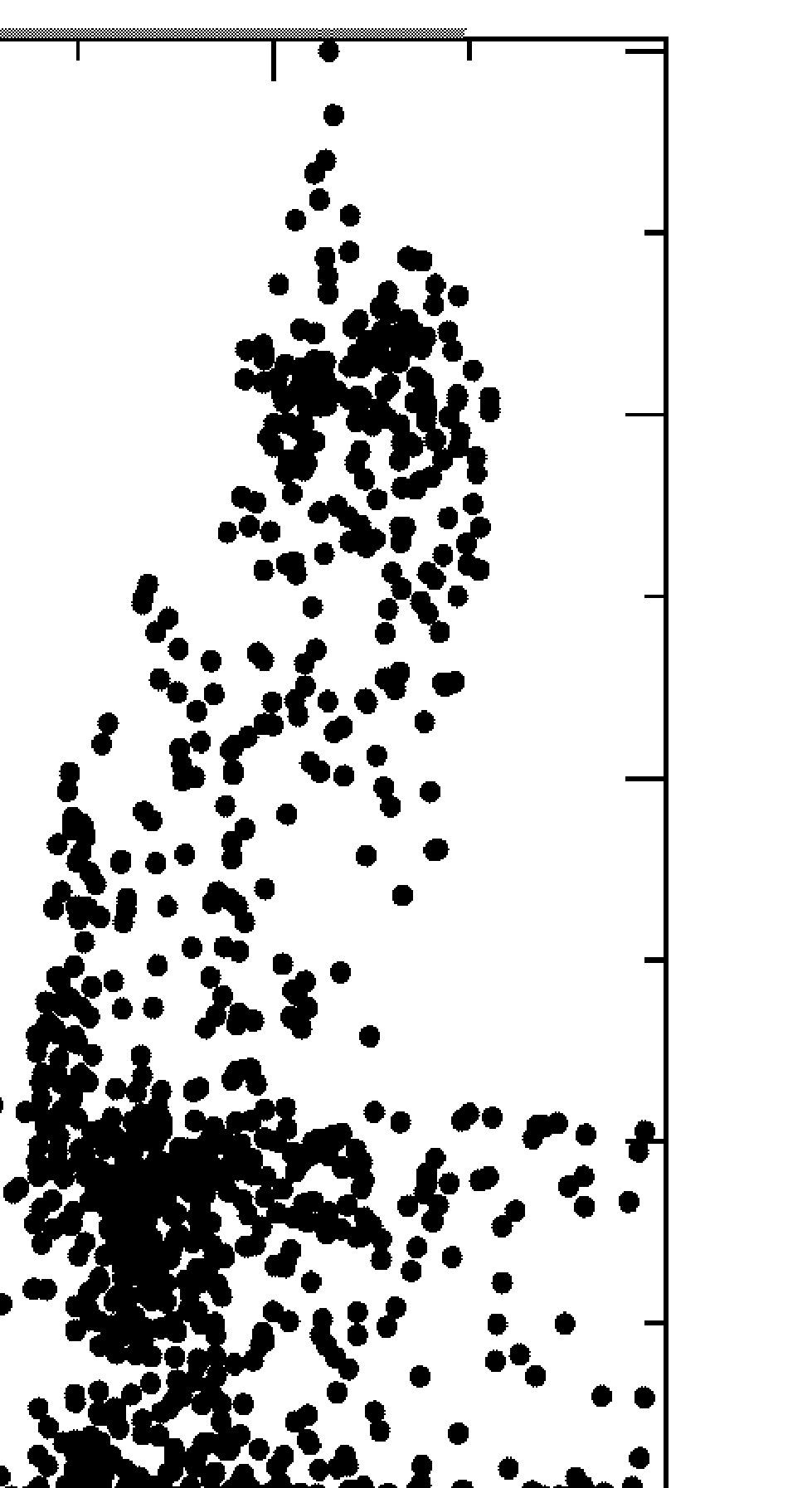}}
        \caption{
 Spontaneous separation into ground-based and attractor based 
regions.  All amoebas are identical, so sorting is not due to differentiation.
\label{fig:3}       
}\end{figure}

\begin{figure}[htb]
\centerline{
\includegraphics[height=8cm]{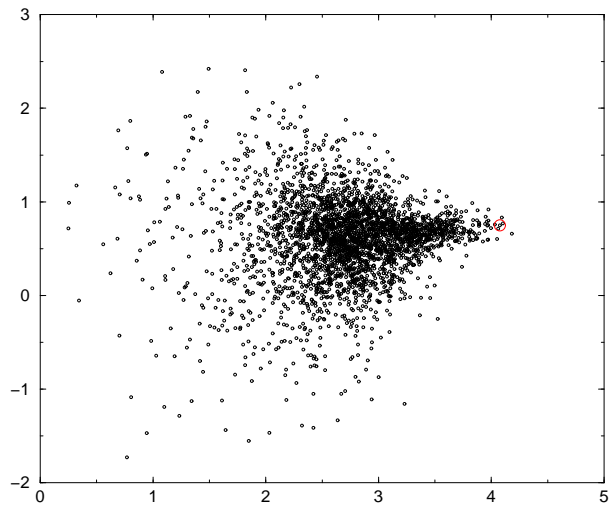}}
        \caption{
 Amoeba positions projected into the horizontal plane as the tower is 
falling.  Amoebas are shown by circles, the red circle surrounds the signaller.  The broad base of the tower forms the broad region to the centre, while the top of 
the tower has fallen to the right. 
\label{fig:4}       
}\end{figure}

The addition of signalling
introduces qualitatively different behaviour.  Assuming the signalling
amoeba is located somewhere near the centre, the other amoebas stream inward
radially, or spirally, if the central amoeba is moving (see figure 1).
This motion generates an external pressure on the resulting colony
which increases as the amoebas become more tightly packed.  Eventually
an instability in the third dimension occurs, and the centre of the
colony is lifted up.  The inward pressure continues and a tower is
formed.  The signalling amoeba is toward the top of the tower, since it
was at the centre of inward motion and therefore near the point of
highest pressure where the instability first sets in, it will be one
of the first (not necessarily the first) to be uplifted (fig 2).

The next
phase depends on the number or particles present and the strength of
the interamoeba potential.  With a few hundred particles the tower
of amoebas is stable, with the signalling amoeba ultimately gathering a
surrounding blob of amoebas atop a thinner stem (see figure 3).  With
more particles (a few thousand), the tower grows taller and thinner
until ultimately it becomes unstable against toppling over.  The
toppling instability occurs while the signaller is still near the top
(Fig 4), and so the fallen tower has self-organised into a slug shape
with the signaller at its head.  The amoebas continue to move towards
the signaller, pushing it forward and the slug moves off at high
coordinated velocity (Fig 5).  While some amoebas can get left behind,
the slug maintains its multicellular integrity, with the signaller
toward the front almost indefinitely. 

 Thus the effect of signalling
is to produce three structures which would not occur for "inert"
amoeba particles: the stem-and-ball, the tower and the slug

\begin{figure}[htb] 
\centerline{
\includegraphics[height=8cm]{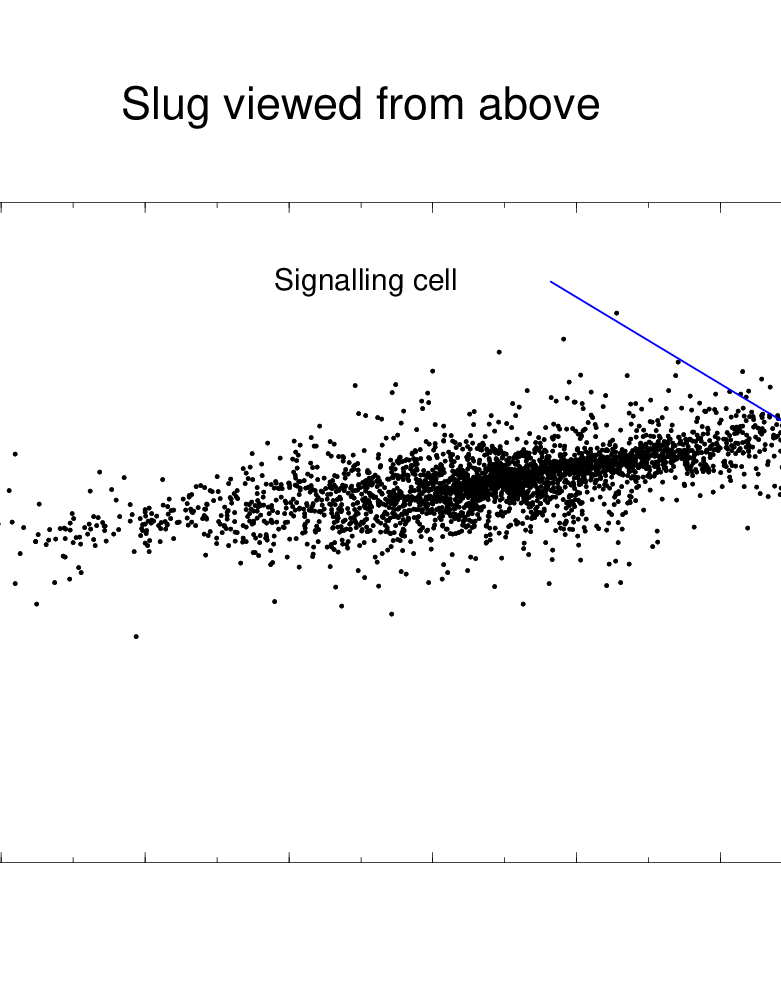}}
        \caption{
Amoeba positions of fully assembled slug in motion projected onto the 
horizontal plane.  Signaller again denoted by red circle.  Thanks to the 
coordinated motion, the slug is able to move much faster than a single amoeba. 
Note the diffuse halo of amoebas on the periphery of the slug: motion would be 
faster still were they more strongly attached to the body, suggesting an 
incremental gain from evolution of slime production. }\label{fig:5}       
\end{figure}

\subsection{Relevance to Dictyostelium
Discoideum} 
\label{sec:5}
The shapes predicted by our simple model are remarkably
similar to those observed in the slime molds\cite{bonbook}.  The slime mold has
fruiting bodies, towers and slugs, of similar general shape to what
the model predicts.  In fact, real mold shapes are more pronounced
than the self assembling shapes we observe, and these shapes are
supported by differentiation of amoebas for specialist tasks (stalk and
ball) and enhanced by slime excretion.  It is interesting, however,
that these basic shapes are just those which can self-assemble in
response to a simple "come hither" signal.  It is plausible that a
primitive single amoeba organism, developing the signalling mechanism,
found the self-assembling multicellular shapes evolutionarily
advantageous, perhaps in foraging (using the coordinated movement of
the slug) or in dispersal (from the fruiting body).  If so,
adaptations such as differentiation and slime emission which enhance
these basic, advantageous shapes could accrue incrementally.  It is
possible that a suitable genetic analysis might establish whether
signalling evolved before slime emission and slime emission before
stalk amoeba differentiation. {\it It is also possible that primitive
organisms still exist which do not excrete slime and, though ancestral
to the slime molds, are not classified with them.}

  We note that the
adaptation to signalling does not confer any advantage to the
signalling amoeba in the absence of other amoebas. Shape formation may
confer advantages to the colony as a whole, and this group advantage
is feeds back as a marginal advantage to the signaller (along with all
others in the colony).

\section{Previous Dictyostelium Models }
\label{sec:6}
A number of previous authors have simulated dictyostelium, and it is
interesting to compare our work with them.  An important distinction
should be drawn; other models assume the feature of modern
dictyostelium, and are thus more complex and specific to the
system. Much of the modelling is devoted to streaming and spiral
formation in the aggregation stage\cite{cohen1,cohen2,dall,McN,Pals,Lev1,Lev2}.
Our model captures the streaming and cAMP spiral formation with 
distinct CA amoeba and without requiring autocyclers.
 Also, many of the previous studies
were in 2D: a feature of this work is that the self-assembly of
the slug is necessarily 3D.  Most notably Palsson (2001) simulated 2D
slugs in motion and the aggregation process in 3D\cite{Pals2,Pals3}.  As in the present
case, they report mound formation without upward chemotaxis, although
it appears that in their model the mound formation is driven by
reducing surface tension, as do fluid-based models\cite{bret,Vasi} 
rather than the
dynamical instability which gives rise to our towers (Fig
1). Contrariwise, in the model of Levine et al (1997)\cite{Lev2} the three
dimensional behavior arises from a probabalistic climbing mechanism:
broadly, each amoeba climbs rather than being pushed up.  The model of
slug motion arising here, of all amoebas pushing forward, is broadly
similar to that of Dormann and Weijer (1997, 2002)\cite{dorm,dorm2}.

\section{Conclusions}
\label{sec:7}
The results obtained using the CA-DE model show a strong similarity to
the real Dictyostelium Discoideum, reproducing the streaming behaviour
and the spiral patterns during aggregation. The model was then used
to examine how the streams form and why the amoeba move faster when in
a group than travelling individually. Finally, a result which has not
been previously seen, the model was used to show that auto-cycling
amoeba are not necessary to trigger the aggregation process.

The full CA-DE model is computationally expensive and could not be extended 
into three dimensions for the timescales needed to obtain mound and slug formation.
Therefore, another simple model has been presented to examine the shapes which can form
in response to a simple signal emitted by one amoeba which attracts
others.  We find that ball-stalk shapes and multiamoeba slugs
self-assemble in response to signalling by a single amoeba. In view of
the simplicity of our model, it is likely that these simple shapes
would be typical of primitive self-assembling organisms, and indeed
they do resemble the multiamoeba components of modern slime molds,
albeit without slime.  

Our model is less complex than previous work,
and thus less directly applicable to the details of each aspect of the
life cycle of modern slime molds.  It does nevertheless capture the
streaming, tower building, slug self assembly and slug motion
exhibited by these systems in a single simple model.  It also suggests
a new picture for the mound/tower formation stage, based on dynamic
instability due to pressure of incoming amoebas, rather than deliberate
climbing or surface tension.  

In view of the model simplicity, and
lack of any dictyostelium-specific input, we believe we have shown
that the shapes and behaviour of dictyostelium, far from being strange
and complex, are just what one would expect from a simple,
self-assembling system.

\end{document}